# Single-molecule kinetics in living cells

**Johan Elf and Irmeli Barkefors**

Department of Cell and Molecular Biology, Uppsala University, 75124, Uppsala, Sweden.

## Abstract

**In the past decades, advances in microscopy have made it possible to study the dynamics of individual biomolecules in vitro and resolve intramolecular kinetics that would otherwise be hidden in ensemble averages. More recently, single-molecule methods have been used to image, localize and track individually labeled macromolecules in the cytoplasm of living cells, allowing investigations of intermolecular kinetics under physiologically relevant conditions. In this review, we illuminate the particular advantages of single-molecule techniques when studying kinetics in living cells and discuss solutions to specific challenges associated with these methods.**

## Introduction

Historically, biochemistry has been performed in vitro and measurements have been averages over innumerable molecules in a test tube. During the last decades, we have refined our scope, and we can now study biochemical reactions as they happen, inside the living organism, one molecule at a time. This allows us to address questions that were previously out of reach because the answer was either masked by ensemble averages or dependent on aspects of molecular biology that cannot be reconstituted in vitro. In the details of intracellular spatial and temporal dynamics, life emerges from the physics and chemistry of dead molecules; single-molecule techniques provide the necessary sensitivity and resolution to watch this mystery unfold in space and time.

In this review, we highlight the particular advantages of single-molecule (SM) techniques when studying dynamics in living cells and present smart solutions that scientists have come up with to solve or circumvent the specific challenges associated with these methods. We do not review the biological findings as such, but rather the specific aspects of the single-molecule toolbox that enables us to study life at the single-molecule level.

The field of single-molecule research has been covered in many excellent recent reviews, which is why we start by demarcating which topics that we will not revisit here. Single-molecule biophysics was until the last decade mainly used to investigate properties of molecules in vitro; in particular, to dissect the asynchronous intramolecular state transitions of molecules that cannot be seen in the enable average. These techniques have allowed us to study the mechanics of molecular machines at exceptional precision, and the development has been reviewed for example in(1, 2).

A large proportion of the recent development in single-molecule detection techniques can be attributed to the super-resolution imaging revolution that needs no further description at this point. Readers that are interested in the development of super-resolution microscopy are referred to these reviews(3–5).

For live-cell applications, there are limitations as to which of the SM techniques that can be used. The challenges are mainly associated with cellular background fluorescence, problems of getting labeled molecules across the intact plasma membrane, and localization issues due to the inherent movement of the living structures. These issues are generally easier to address when working with membrane-bound molecules. The cellular background fluorescence can be minimized by the use of total internal reflection fluorescence (TIRF) microscopy, and the molecules on the outer membrane can be labeled externally, circumventing the problem of membrane impermeability. Diffusion is also generally slower in the membrane, which means that it is relatively easy to get good signals. Thus, membrane-bound single-molecule kinetics is in itself a big field that has been reviewed elsewhere(6, 7).

What remains within the scope of this review are the techniques that can be used to detect single molecules in the cytoplasm of living cells. Much like other molecular techniques, like PCR or DNA sequencing, SM imaging and tracking can be varied in many ways. We present examples of different technical flavors, and it is up to the reader to inter- or extrapolate among these to make the measurements needed for each specific application. That said, many vital concepts are reoccurring. We focus on shared challenges and give examples of how they have been overcome in specific studies. The section on analysis is necessarily biased to our specific interests as there are nearly as many analysis methods as there are experiments or at least experimentalists. However, before we dive into the technicalities, we start by describing three general use cases for dynamic single molecule methods in living cells (Fig. 1).

## 1. Counting and cell-to-cell heterogeneity

Single-molecule sensitive methods are needed to determine the number of a particular molecular species and how this number varies over time, in space or between cells. The use of single-molecule counting falls into a few different types of experiments.

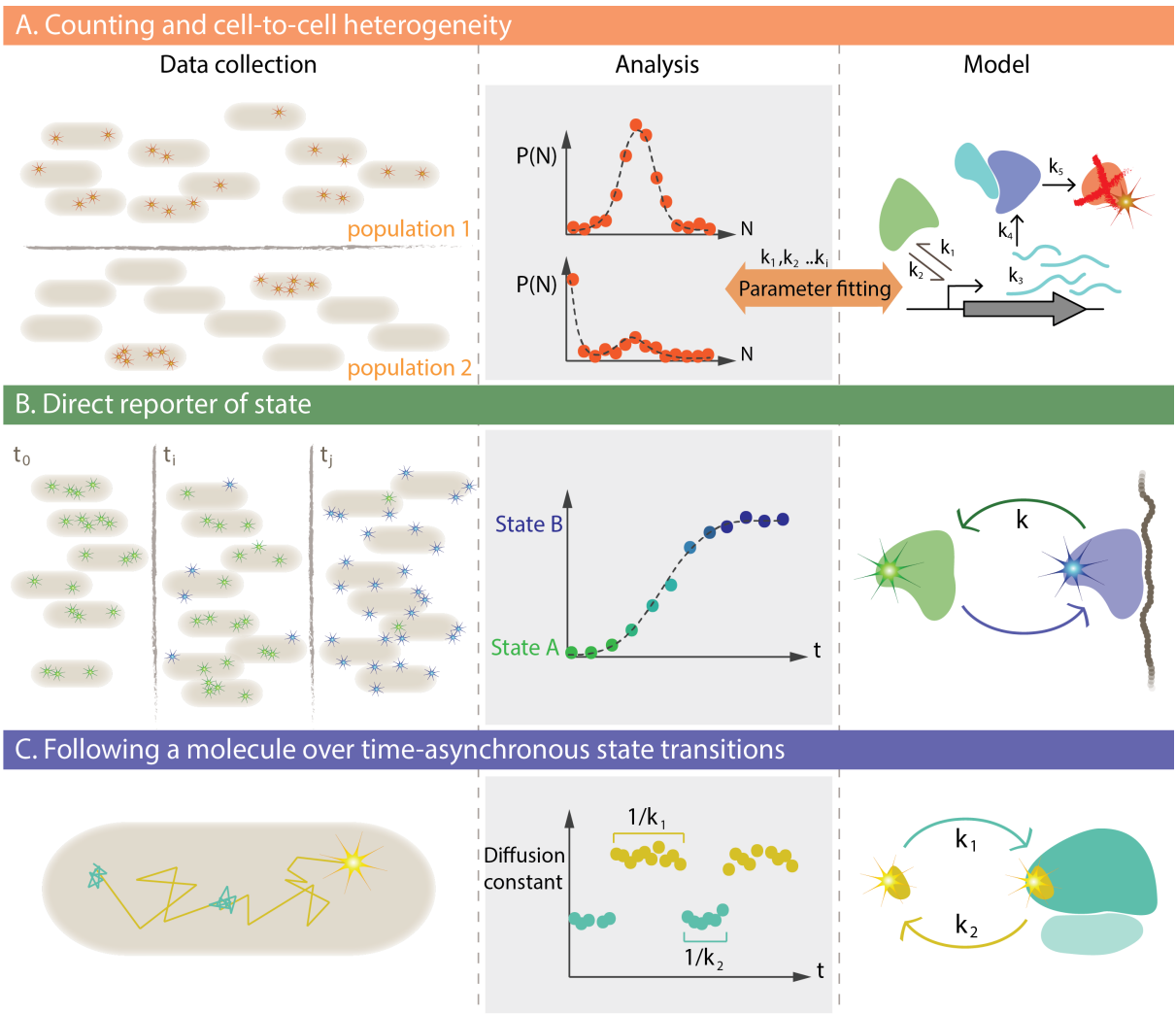

*Figure 1. **Different approaches to obtaining kinetic information from single-molecule experiments in living cells.** A. The steady-state copy number distributions within a cell population are sometimes possible to fit with stochastic models including unknown parameters. B. The fraction of molecules in various states of binding or diffusion can be followed in time after a perturbation at $t_0$. C. The individual molecules' state transitions can be followed directly to obtain intracellular dwell times and rates.*

**Copy numbers**

An apparent single-molecule application is to count molecules based on how many fluorescent spots that are observed(8). The major challenges associated with this approach are the stoichiometry in labeling and the maturation of the fluorophore. This approach also fails when there are many molecules such that the spots are overlapping, in which case the total number of molecules can be estimated by the total fluorescent signal integrated over the cell, divided by the average fluorescent signal for an individual molecule(9). This method is only an approximation to the number of molecules in an individual cell for many reasons. For example, the average fluorescence of an individual molecule is estimated for a non-blinked out fluorophore whereas the total integrated fluorescence includes both dark and bright fluorophores. It is also hard to make the correct background fluorescence subtraction.

It is sometimes possible to count the number of bleaching steps to determine the stoichiometry of fluorescent molecules in complexes, assuming that the individual fluorophores are sufficiently bright(10, 11). Again, it is essential to consider that only a fraction of the fluorophores will be mature in a live-cell situation(12).

**Inference of kinetics from steady-state distributions**
An obvious strength of single-molecule counting in cells is the possibility to characterize the cell-to-cell variation in molecule numbers and determine the steady-state distribution from snapshots over different cells. These distributions arise from stochasticity in the underlying processes, such as gene expression, fluorophore maturation, and protein partitioning in cell division, and depend on the mechanisms and kinetic rates specific for each process(13). By assuming a dynamic model, it might be possible to fit the model parameters based on the steady-state distribution(14–18) and thus estimate the underlying kinetic rates (Fig. 1A). When using such methods, it is important to keep in mind that different stochastic models can give rise to very similar steady-state distributions(19) and that deterministic differences between individual cells can also be mistaken for stochastic effects(20, 21).

**Following actual low copy dynamics**
Following individual cells over time makes it possible to study dynamic correlations that cannot be seen in steady-state distributions. Some classical examples are the early single-molecule studies of bursts in protein(22, 23) or RNA(24) expression from the lac operon. Looking at numbers of molecules in single cells over time, it is also possible to study how low copy number dynamics give rise to phenotypic copy number transitions. For example, Choi et al.(25) monitored fluctuations in the expression of fluorescently labeled lacY permeases to deduce how many LacY that are needed to switch the bistability observed for the Lac operon in the presence of Methyl-β-D-thiogalactoside (TMG). Similar in nature is the work by Uphoff et al., which shows that the DNA methylation repair by Ada is turned on by a single expression event which then activates further expression of the protein by positive feedback(26).

## 2. Direct reporter of state

In some cases, the single-molecule readout can provide useful information about the state of a molecular species in the cell. For example, we can measure the fractions of mobile and immobile molecules, how these fractions are distributed in the cell and whether a particular binding site is occupied or not. By monitoring how the occupancy of specific molecular states changes over time following a perturbation, it is possible to obtain kinetic information without following individual molecules through several state transitions. These experiments can in principle be performed in bulk since they are synchronized by a perturbation, but single-molecule detection is still needed to detect the binding states or spatial localization with sufficient time resolution.

**Binding states**

A typical study of this kind measures what fraction of labeled molecules that is bound to a largely immobile structure, e.g., a fluorescent transcription factor binding to the chromosome(27), as a function of time after a perturbation (Fig. 1B). Although transcription factor dissociation from a chromosomal operator upon a chemical cue could in principle be monitored by a gene expression assay, the time resolution is significantly improved if the binding state of the operator site is measured directly. In this category, we find the transcription factor binding studies by Hammar et al.(27, 28). In the first study, the transcription factor LacI's association was triggered by a chemical cue, after which the fraction of LacI that was bound to the *lacO*-operator was measured as a function of time. There is only one *lacO*-operator per chromosome, which makes single-molecule detection particularly advantageous in this case. A similar assay was used in the follow-up study to measure transcription factor dissociation by chasing bound fluorescent molecules with non-fluorescent molecules to prevent rebinding(28). This approach was also used to measure Cas9 search kinetics by programming fluorescent dCas9 to bind lacO-operators made available by addition of Isopropyl β-D-1-thiogalactopyranoside (IPTG)(9).

**Movement states**

Another type of mobility-based assay determines the diffusive state of a molecule by single-molecule tracking and uses this information to deduce the typical location of molecules in a particular diffusive state. With this type of assay, different groups have studied the spatial distribution of translating ribosomes versus free ribosomal subunits in E. coli(29, 30). Similar strategies have also been used to study RNA polymerase activity(31, 32), transcription factor search strategies in mammalian cells(33)(34), tRNA interactions with the ribosome in bacteria(35), and the different modes of action of the bacterial SMC (structural maintenance of chromosomes) complex(36). Methodologically, this type of kinetic measurement was pioneered by English et al.(37) who showed that the mobility of the RelA protein increased in response to amino acid starvation. The RelA response described in this study has however been hard to reproduce. Recent studies(38, 39) report decreased RelA mobility in response to starvation, and the function of the fluorescent fusion protein used in(37) has been questioned as current cryo-EM studies put the label inside the ribosome(40).

# 3. Following a molecule over time - Asynchronous state transitions

Intracellular processes can rarely be synchronized, which is why it makes sense to determine transition rates by monitoring state changes one molecule at a time. Such single molecule measurements in vivo are restricted to binding and dissociation reactions that can be explored using single-molecule tracking to reveal changes in translational or rotational diffusion, spatial position, polarization or FRET (Fig. 1C).

Even without following a molecule through different states of binding, it is possible to determine the dwell times in bound, non-diffusive, states by varying the exposure time and compensating with laser power to determine on which timescale the molecules are immobile(9, 42). It is also possible to use a constant exposure time and augment the interframe time to look for timescales where the molecule is observed in consecutive frames(11, 34, 43). Single-molecule dwell time distributions were for example used to elucidate the kinetics of transcription regulator FoxF1(44).

Contributions where kinetic properties are learned by following individual molecules in time include studies of DNA mismatch repair by MutS(45), enhanceosome assembly(46), Hfq-RNA binding dynamics(47), FtsZ thread milling(48, 49), P53 dynamics(50) and tRNA translation speed(51). A significant challenge in these systems is that bleaching and blinking of the fluorophores limit the lengths of trajectories that can be recorded from a single molecule. Depending on the quality of the label and the intracellular background fluorescence it is usually possible to record around 10 to 20 frames from each fluorescent protein molecule, which will limit how many state transitions that can be captured. This problem can be solved to some extent by model-based data analysis techniques. By correctly combining the information from thousands of relatively short trajectories, it is possible to learn model parameters, such as transition rates and the number of diffusive states, from the experimental data at the cost of assuming a kinetic model.

This class of SM experiments also includes the more traditional in vitro single-molecule experiments studying the intramolecular state transitions of molecules using, for example, SM-FRET(41), single-molecule enzymology, optical traps, and multicolor colocalization studies of complex assembly. Many of these experiments would undoubtedly be very informative if executed in living cells and although the level of difficulty is in most cases staggering, some progress has been made with electroporated FRET probes and fluorescent oligonucleotides(51–54).

# Labeling

Labeling is typically the most challenging aspect of any single-molecule experiment inside living cells. In order of importance, labels should be (1) specific (2) non-perturbing (3) stoichiometric, and usually also (4) bright, non-blinking, yellow to red and bleach resistant.

Due to the high signal-to-background ratio of the fluorescence signal given the size of the label, virtually all single-molecule experiments in living cells depend on fluorescence labeling. Examples of in vivo single-molecule studies that have not relied on fluorescent labels have exploited luminescence(55, 56), fluorogenic substrates(23), gold nanoparticles(57, 58) or interferometric scattering microscopy (iSCAT)(59, 60).

**Fluorescent fusion proteins**
The primary challenge of intracellular labeling is specificity; it is crucial that the label be on the molecule of interest and not elsewhere. For proteins, the most commonly used strategy to accomplish specificity is to fuse the target to a fluorescent protein. In this way, genetically directed specificity is achieved by making the fusion protein in one translation unit. Fluorescent fusion proteins have been the workhorse of live-cell single-molecule studies including the early work on lamellipodia(61), cell division(62), membrane dynamics(63) and gene expression(22).

Apart from their specificity, FPs fall short in essentially all aspects. They are large, around 20 kDa, often as large as the protein of interest which implies that impaired activity of the fusion protein is a concern. They have notoriously bad photophysics, i.e., they blink irregularly(64, 65) and bleach easily which usually results in short trajectories(8), and generally need a long time to mature before they are detectable(12). FPs also tend to form aggregates which, apart from impairing function, might also cause miscounting and incorrect localization(66). As an example, the bacterial actin homolog MreB was observed to form cell spanning helices, which were later discovered to be artifacts caused by the N-terminal fluorescent fusion. Instead of long helices, MreB forms short, dynamic filament bundles that treadmill around the cell body(67, 68).

If counting expressed proteins at the single-molecule level is the primary goal, a clever trick to overcome some of the problems with misbehaving fusion proteins was introduced by Xiao and colleagues(69). By introducing a cleavable linker between the protein of interest and the FP, it is possible to get stoichiometric expression of the FP without impairing protein function. Counting of low copy number molecules in the cytoplasm of prokaryotes can also be facilitated by mechanically compressing the cells to reduce background and slow down diffusion. Okumus et al. use a microfluidic approach to compress E. coli and perform high throughput single-molecule counting of low copy number cytoplasmic FPs(70).

The SunTag technology(71) can be useful as a single-molecule reporter since the complex is long-lived and bright. The signal amplification of the SunTag is based on detection of multiple scFV-sfGFPs binding to a single genetically encoded reporter protein. Although its bulky size makes the SunTag suboptimal for functional studies of single proteins, it can be useful for larger structures as demonstrated by Liu et al., who used the technique to study synaptic vesicle dynamics in cultured neurons and intact zebrafish(72).

If the objective is following individual molecules in time, blinking and bleaching become severe limitations. An important method to get more, albeit not longer, trajectories from single-molecule tracking experiments is to use sptPALM(73, 74). The technique makes use of photoconvertible FPs that are activated one or a few at a time, making it possible to record many trajectories from one single cell. In the analysis section, we describe the efforts that have been made to extract meaningful data from many short trajectories using statistical learning methods. sptPALM is also a powerful tool to map the averaged local diffusion properties of a molecule of interest in individual cells(29, 75–78). However, a specific problem arises when the technique is used to study the binding of single proteins to a limited number of binding sites since these few binding positions are likely blocked by non-converted molecules.

Many attempts have been made to increase the length of SPT trajectories by improving the photophysical properties of genetic fusions, mainly by modifying existing FPs(12). Examples of FPs that have been successfully used in several SM tracking studies are mEOS2(43), PA-mCherry(11) and Dendra2(33, 37, 79). Improved FPs are engineered continuously, and any attempt at a comprehensive list would be inconclusive before this review is in print.

If time resolution is not an issue, the absolute length of trajectories can be increased by spacing out the observations as exemplified by Ho et al. who measured residence times of several minutes for a Mfd-Ypet fusion protein(80). When making sparse observations, there is always the possibility that dissociation and rebinding happen between frames, which makes pure residence time measurements complicated. This problem can be circumvented by chasing the fluorescence molecules with a high concentration of non-fluorescent molecules to prevent rebinding of fluorophores after dissociation(28).

Finally, split fluorescent proteins can be used as a tool to probe the location and dynamics of individual interacting protein pairs. The technique was successfully demonstrated by Liu et al. who used bimolecular fluorescence complementation (BiFC)-PALM to study the interaction of MreB and EF-Tu in E. coli(81).

**Labeling proteins with organic dyes**

When we look beyond intrinsically fluorescent proteins, we have to sacrifice some of their excellent chemical specificity for improved photophysics. One avenue that has been successful for SM studies is the use of genetically encoded protein adaptors that specifically bind a dye molecule that is externally supplied. For example, HALO(82) and SNAP(83) tags bound to Tetramethylrhodamine (TMR) dyes have been used to track cytoplasmic proteins in eukaryotes. Zhao et al. used the blinking photophysics of the TMR dyes to count individual RNA polymerases(84) and Chen et al. investigated the search kinetics of the transcription factors Sox2 and Oct4(46). The technique has also been applied in bacterial systems(85) where membrane impermeability of the dye is usually a problem. Barlag et al. tracked individual proteins fused to the Halo or SNAP tags in Salmonella enterica(86), and a similar set-up makes use of the affinity of the E. coli dihydrofolate reductase (eDHFR) for trimethoprim (TMP) photoconvertible dye to allow dynamic imaging of single histones by dSTORM(87). More recently, excellent dyes with good membrane permeability like Silicon-Rhodamines(88) and JF dyes(89–91) have been used for single-molecule live-cell tracking applications(48, 92–94).

Small organic dyes have exceptional brightness and bleach resistance, but membrane permeability of the dye, as well as nonspecific sticking(95), are critical problems that need to be carefully controlled for, for example by running the experimental protocols without the target. The SNAP and HALO tags are also almost as large as FPs, with the associated risk of impaired protein function, and ongoing research efforts aim to develop smaller peptide tags that can be reacted specifically with dyes(96). One of the more promising SM applications along this line is the dL5 peptide which activates Malachite Green and is reported to give excellent brightness and bleach stability(97). In a recent study, Virant et al.(98) tracked diffusing molecules in live HeLa cells using dSTORM and a peptide tag-specific fluorescent(AF647) nanobody. With a molecular weight of only 12–15 kDa, the nanobody is considerably smaller than, e.g., GFP or the HALO-tag.

The ultimate small protein label is a single amino acid, and the recent progress in genetically encoded unnatural amino acids(99, 100), in combination with recoded organisms(101), suggest that it is possible to make specific single amino acid labels in living cells. Cheng et al. used labeling of an unnatural amino acid to study the dynamics of the growth factor type II receptor T(beta)RII on the surface of HeLe calls(102). In an attempt to use the technique to track proteins in the cytoplasm, Kipper et al.(103) introduced the orthogonal pyrolysin aminoacyl tRNA synthesis system into a recoded E. coli with the Amber stop codon in a selected position in the protein of interest. The Amber codon is translated into the clickable TCO*-AK amino acid residue which is reacted with 1,2,4,5-tetrazine fluorophores in the living cell. The scheme works nicely for single-molecule tracking applications in the outer membrane, but background from non-clicked dye limits the intracellular applications so far. A possible way forward would be to combine labeling of unnatural amino acids with turn-on dyes, which are non-fluorescent until they are clicked with the target(104).

For eukaryotic cells, microinjection of pre-labeled proteins is a good option to intracellular labeling. It gives full freedom in labeling sites and dyes at the cost of not being able to run the experiment over several cell generations. For example, the Dahan lab has successfully injected and traced Atto647N labeled TetR in human cell nuclei to study its search kinetics(34).

smFRET(105) has been instrumental for studying inter- and intramolecular dynamics in vitro and during the past decade the technique has moved into living cells, enabling detection of the in vivo conformation of individual proteins at very high time resolution. Sakon and Weninger microinjected site-specifically labeled recombinant SNARE proteins with a FRET donor and acceptor into cultured eukaryotic cells and detected the conformational changes as the SNARES engaged in complex formation(106). Köning et al. used smFRET to study conformational dynamics and folding of different proteins in the cytoplasm of cultured HeLa cells(107).

In addition to the general labeling strategies, there are more specific stains, FRET probes and biosensors with great potential for single-molecule studies. One noteworthy example is FLINC biosensors(108) that make use of the proximity-based blinking of TagRFP and Dronpa to monitor the spatial activity of enzymes at super-resolution length scales using Super-resolution Optical Fluctuation Imaging (SOFI)(109, 110).

**Labeling proteins with Inorganic dyes**
Inorganic fluorescent nanoparticles like quantum dots (QDs) and other nanocrystals are even brighter and more stable than organic fluorescent dyes, but they are quite large, difficult to get across the plasma membrane and cytotoxic unless adequately functionalized. Nevertheless, quantum dots have been used successfully to track proteins and virus particles in living cells. The progress that has been made using nanoparticles for single-molecule tracking was recently reviewed by Jin et al.(111).

**Labeling nucleic acids**
In the absence of inherently fluorescent RNA, progress in this area has primarily been made with fluorescent proteins fused to RNA phage coat protein motifs. In particular, proteins from bacteriophages MS2(112) and PP7 systems(113) fit into successful single-molecule applications, where arrays of binding sites enable signal above the background of non-bound MS2-GFP(114, 115).

By monitoring the gradual increase and abrupt decrease in fluorescence from transcription events, Larson et al. measured the kinetics of transcription elongation in real time at individual loci(116). The Singer group has also used the MS2 and PP7 system in parallel, to study independent mRNA expression from two different

alleles(115), and by exploiting fluorescence complementation, they manage to remove background fluorescence almost completely(117). The group also studies translational hot-spots by simultaneous tracking of mRNA and ribosomes(118). Morisaki et al.(119) use a combination of epitope labeling and MS2 fusion to study translation kinetics by tracking the nascent chain and Wu et al. also studies translation kinetics by a similar approach using the SUNTag to boast the fluorescent signal(120).

However, the MS2 array is not an optimal tool to determine the instantaneous number of mRNA per cell since the nucleoprotein complexes have been immortalized, preventing degradation of the labeled mRNAs(24, 121). This problem was addressed in a recent paper by the Singer lab where they describe a reporter system with reduced affinity for the MS2 coat protein, allowing mRNA degradation while still producing signals that are strong enough for single-molecule detection(121).

In want of intrinsically fluorescent RNA for direct intracellular labeling for single-molecule studies, the hope has been to evolve an RNA aptamer capable of binding a fluorogenic cofactor. Spinach varieties have been considered promising candidates but are not yet sufficiently bright(122). A seemingly more promising route is to evolve an aptamer that binds a fluorescence quencher that is provided bound to a fluorophore. Binding strengths and turn-on ratios are approaching the regime that could make this strategy useful for single-molecule work(123).

Tiny Molecular Beacons(124, 125) have frequently been used to access mRNA dynamics in living cells, but the method has been hampered by problems with nonspecific binding and false positives. Zhao et al.(126) demonstrated how loop-domain phosphorothioate modification can be used to reduce nonspecific signals, making molecular beacons a viable option for sensitive and accurate imaging of individual RNAs at the single-molecule level.

Electroporation of labeled tRNA(35) has been successfully used to study tRNA localization inside the cell(54) and to monitor translation rates at the single codon level(51). Studies of intracellular mechanisms by transfecting RNA or DNA labeled with organic fluorophores have also been shown to work in the eukaryotic model system S. Cerevisiae(52). For larger cells, microinjection is an alternative to transfection as shown by e.g. Bratu et al.(124) and Pitchiaya et al.(127).

Finally, there have been successful attempts at studying the structural conformation of DNA in vivo by smFRET. For this purpose, DNA constructs labeled with FRET donor/acceptor pairs have been introduced in E. coli by heat shock(128) or by electroporation(52).

# Imaging

**The basics**

Basic wide-field single-molecule live-cell imaging(22) and tracking(8, 129) are no longer particularly complicated from an instrument perspective. The super-resolution microscopy revolution has paved the way for a number of plug-and-play systems that work well for live-cell single-molecule wide-field epifluorescence detection. Assembling your own systems is also easier than may be expected; most live-cell single-molecule tracking applications use a simple DPSS (Diode-Pumped Solid-State) laser that is focused down at the back focal plane of the objective such that the sample is hit with a collimated beam. Fluorescence is collected by a high numerical aperture objective and separated from the excitation light by an optimal dichroic and emission filter before the image is collected on a very sensitive (>90% quantum efficiency) EMCCD or sCMOS camera. The total cost of the above setup is less than 100,000 USD, and it will do the trick of detecting and tracking individual, slowly moving (< 1 µm$^2$/s), yellow or red fluorophores in bacterial cells at timescales down to 5ms.

A fundamental challenge with wide-field epifluorescence imaging is reducing the background autofluorescence. Living cells emit significant fluorescence when excited below 500 nm, which implies that the fluorophore should ideally be yellow or red. Using a laser line that is just below the excitation maximum reduces the autofluorescence per excitation event.

**Thick samples**

For samples thicker than a few micrometers, simple wide-field epifluorescence is usually not sufficient for single-molecule detection. The light excites the whole depth of the sample and fluorescence is detected from out of focus sources drowning the signal of interest. In conventional in vitro SM imaging where the sample is at the cover glass surface, the problem is usually solved by total internal reflection fluorescence (TIRF) microscopy where only fluorophores within 200 nm of the glass surface are illuminated. A similar effect can be achieved by HILO(130), sheet illumination(43, 131), aberration corrected multifocus microscopy(132) or Bessel beam illumination(133, 134), to study objects that are further away from the focal plane. For example, Izeddin et al. used HILO to track single transcription factors in eukaryotic nuclei(33). Eukaryotic transcription factors were also tracked using the extension of the Bessel beam to lattice light sheet microscopy(46, 135, 136).

**Stroboscopic excitation**

In tracking applications, the excitation time can be modified with the fundamental principle that the exposure time in each frame should be short enough to avoid motion blur. For example, if a movement of 50nm is tolerated during each exposure, the exposure time needs to be in the order of 1ms, i.e. (0.05um)^2 /(2x1um^2/s), if the molecule is moving at 1um^2/s. Many cameras do not operate at 1ms per frame

(1 kHz) for large regions of interests, why it might be favorable to use stroboscopic laser excitation instead of fast exposure(8). Stroboscopic illumination reduces the excitation time by triggering the laser for a short period in the middle of the camera exposure time, thereby minimizing the motion blur. The technique can be quite easily implemented with a signal generator that triggers the laser, or a fast shutter such as an AOTF; based on a TTL signal from the camera.

If the molecules alternate between largely immobile and rapidly diffusing states, it is possible to filter out the immobile molecules by choosing an appropriate exposure time and laser power(8). This strategy can also be used to infer the distribution of residence times from the fraction of molecules that are observed at different times scales(9, 42), a method we refer to as VIBE (Variable Intensity Balancing Excitation)

**Creating a flat excitation profile**
Conventionally, the Gaussian excitation beam is expanded before the microscope such that the essentially flat center is selected to hit the sample while the rest of the beam is blocked out to avoid unnecessary background excitation and auto-fluorescence. This means that much of the laser power is sacrificed in order to achieve a reasonably flat excitation profile. To overcome this problem, the Manley lab has introduced a low-cost microlens array (MLA)-based epi-illumination system. The FIFI (Flat Illumination for Field-Independent imaging) technology increases the throughput of single-molecule localization imaging by capturing larger fields of view with sustained quality(137).

**Going 3D**
On the emission side, tricks can be played with the point spread function of the emitted photons. One of the more popular modifications is to place an astigmatic lens in front of the camera to create an elliptical PSF that stretches in different directions based on the z-position of the sample molecule(138, 139). This method has for example been used to study the dynamics of FtsZ-ring formation in E. coli(140). A similar effect can also be obtained using adaptive optics(141).

The z-range can be much extended with the use of a Double-Helix Point Spread Function (DH-PSF), as generated by a spatial light modulator or a specific phase mask in the emission path(142). As a result, each single molecule appears as two spots on the camera, the orientation of which correlates with the z-position. The DH-PFS has for example been used to study the correlated motion of two different chromosomal loci(143). Even more creative ways to modulate the phase of the PSF by introducing a spatial light modulator (SLM) in the pupil plane are reviewed in(144). One important piece of information that can be obtained from individual emitters is their orientation, which may lead to significant mislocalization of the center of the Gaussian PSF if the emitter is not free to rotate(145, 146). Furthermore, a long-range DH-PFS can advantageously be combined with a dynamic light sheet set-up to lower the background and extend the dynamic range(147), which should have interesting applications in SM tracking where the light-sheet can be adapted to the position of the molecule.

An alternative method to obtain 3D information from a single fluorophore is biplane imaging where two different focal planes are imaged at the same time such that the 3D position can be interpolated based on the shape of the point spread functions. A powerful extension is aberration-corrected multifocus microscopy which has been used to track Halo-tagged RNA polymerase II molecules in U2OS cells(132). Imaging in multiple planes or manipulating the PSF dilutes the photons on the camera chip, which is a critical problem when operating at low signals. A method for accurate 3D localization that does not suffer from this limitation is interferometric (i)PALM(148). iPALM uses two objectives that are mounted in a 4Pi configuration where the photons that are propagated through both objectives are allowed to self-interfere before detection. The optical alignment is complicated and the imaging geometry is challenging for use with live cell samples which may explain the lack of applications in living cells so far, despite the ingenious design. Along the same lines, we may get holographic imaging systems for 3D SMT in the near future assuming the dyes get sufficiently bright(149).

**MINFLUX imaging**
Direct camera-based imaging approaches to record the location of a molecule has a fundamental limitation in the spatio-temporal resolution; hundreds or thousands of photons are needed to pinpoint the location of the molecule to 10s of nanometers since the individual photons are collected in a diffraction-limited range. The problem is twofold. First, it takes time to acquire the photons from the fluorophores, at least 1 ms, which limits the temporal resolution. Secondly, the excessive use of photons exhausts the photon budget, thus limiting the number of frames that it is possible to record.

One way to capitalize more efficiently on the available photon budget is to use a confocal system where the position of the laser focus keeps track of the position of the fluorophore directly. In this way it is possible to deduce whether the position of the molecule and the excitation laser beam coincide just by detecting a few photons. To accurately pinpoint the position, it is necessary to triangulate based on the relative number of photons detected in a few positions, but compared to camera-based detection it is still much more efficient. The challenging part is implementing the feedback system designed to follow the moving molecule with the laser beam(150).

A powerful implementation of the confocal tracking scheme is MINFLUX microscopy(151) where the excitation laser focus can be donut-shaped which means that no photon is emitted when the molecule is in the center of the beam. In this way, it is theoretically possible to know that the laser is in the wrong position as soon as one photon is detected, i.e. no photons are wasted when the molecule is bound and immobile. In reality, the background autofluorescence and blinking properties of the fluorophores make it necessary to use more than one photon. Localization within a few microseconds can however still be made from the number of photons that are typically emitted from conventional dyes and fluorescent proteins, which significantly increases the time resolution compared to camera-based tracking. The other advantage of using a confocal system with a point detector is that polarization data can be recorded for each photon emission event. Polarization detection enables rotational correlation function measurements while tracking the molecule (Elf et al., unpublished).

## Analysis

Making quantitative sense of single-molecule data can be as challenging as obtaining it in the first place. Even with optimal experimental conditions, image analysis can be a challenge due to e.g. low signal to noise ratios, blinking, bleaching and too dense labeling. This can severely limit the information content in the resulting derived data and yield challenging inference problems.

A typical analysis workflow for single-molecule in vivo data includes detecting potential spots, localizing and validating spot candidates and connecting them to individual cells or cellular structures. Tracking experiments also include a particle trajectory building step as well as analysis of position and trajectory data. These tasks are not strictly independent, and by now the collection of approaches, algorithms and software packages are large and diverse. There have been some benchmarking studies on various aspects of imaging and tracking(152, 153) to help with the choice, and more are underway.

In this section, we present an overview of some of the methods available for identifying and localizing spots, classifying them into trajectories, and analyzing the results. The intellectual depth of the analysis methods ranges from entirely ad hoc to statistically rigorous. We also note that the advanced methods are rarely used in high impact biological studies, which suggests a potential for improvement. Alternatively, the most important challenges lie elsewhere, i.e. with the choice of biological question, labeling or microscopy method.

We have no possibility to be comprehensive in this section since the requirements of the analysis method will vary with the biological question in mind. Instead, we focus on camera-based single-molecule tracking starting from a basic pipeline and discuss some possible extensions. A commonly used pipeline(31, 45, 154), is to 1) filter out what might be spots in each image frame using, for example, intensity thresholding or selecting intensity peaks, 2) fit the candidate spots with a Gaussian function and discarding spots that have undesired width or intensities and 3) connect the remaining spots into trajectories if they are within a certain distance in consecutive images. If there are multiple spots within that distance, the sum of displacements is minimized.

Many trajectory analysis methods use mean square displacement (MSD) plots to obtain the diffusivity of individual molecules and to map diffusivity to specific regions in space(29)(Fig. 2A). The slope of the MSD curve gives the diffusion constant at a particular timescale and the curvature holds information about the mode of motion and confinement effects. A straight line over some timescale indicates a normal diffusion in that range. To a first approximation, the intercept of the MSD curve with the y-axis equals two times the localization accuracy of a spot, since this is the movement that would be erroneously observed for an immobile molecule. A statistical difficulty with MSD analysis is that the reusing of data (all trajectory points contribute to all MSD points) can lead to strongly correlated statistical errors, which makes MSD curves challenging to interpret and analyze(155). For a particle undergoing simple diffusion, there are simpler and more robust ways to estimate the diffusion constant(155).

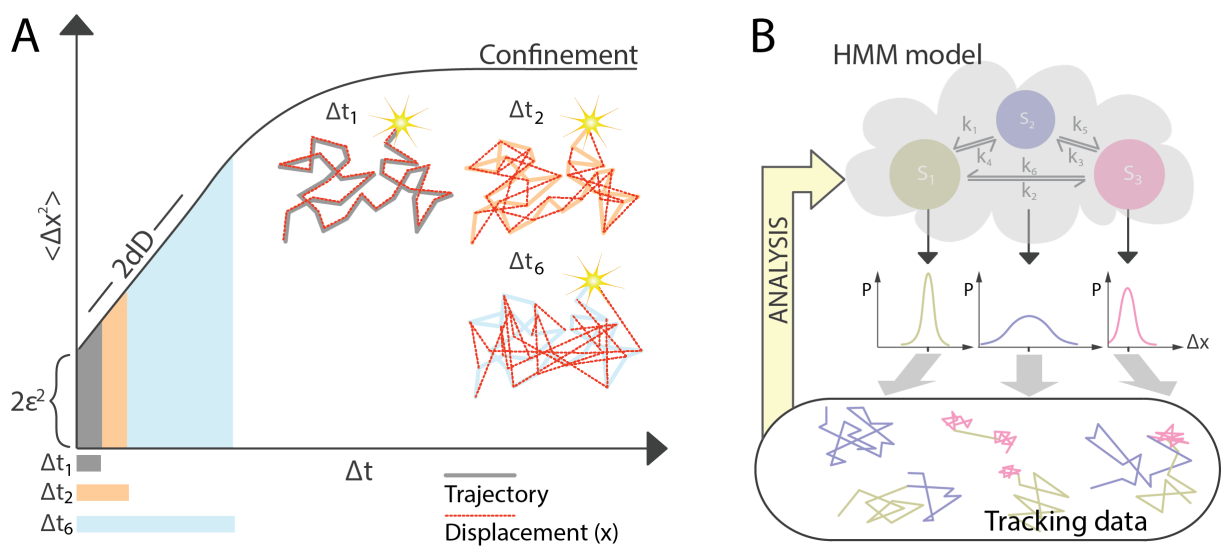

*Figure 2. Particle trajectory analysis. A. Mean square displacement (MSD) is a standard measure of particle diffusion. If a particle displays normal diffusion, the slope of the curve is constant while spatial confinement will result in a plateau. B. In the HMM analysis, the states of the hidden model are defined by different probability distributions from which the e xperimental data are assumed to be sampled. By analyzing the trajectories, the number of states, state occupancies, and transition rates can be recovered.*

**Detecting spots**

The problem of spot detection in noisy images is a classical image analysis problem that has been approached and solved in numerous ways.

A frequently used class of methods is based on the Laplacian of Gaussian filtering(156) where the image is first smoothened by a Gaussian filter of a chosen peak width (σ) after which significant changes in gradients are detected with a Laplacian operation. This procedure makes it possible to find symmetric blobs of a certain size related to sigma. One can also design multiscale blob/spot detection tools following the adaptive space-scale approach(157). This concept can be generalized to local shape as well(158), which performs better for detecting spots deformed by optical aberrations or motion blur. Magnusson et al.(153) consider the image to be a convolution of the image of particles in unknown locations with the point spread function. The particle image is found as an optimization problem with a tunable cost function for the number of particles and a fixed PSF.

An alternative approach, which is well-grounded in theory, is the use of a likelihood-ratio test(159, 160), which fits a Gaussian spot model to the region surrounding each pixel. The method subsequently performs a statistical comparison of the "spot" vs "no spot" hypotheses, with acceptance criteria explicitly set to control the rates of various error types. This method is theoretically attractive since one can, in principle, use realistic models for both spot shape and noise, but computationally demanding since it requires one maximum likelihood fit for every pixel.

A less statistically rigorous but fast and robust method is based on local radial symmetry(161). Spot detection is achieved by determining how each pixel contributes to the symmetry of the neighboring pixels. The method was developed as a facial image detector, but has proven to be very useful also in fluorescence spot detection.

Finally, multi-scale wavelet transform, where the spots of particular sizes are strongly enhanced over the background in selected wavelet planes, is a very fast method that is insensitive to variations in the background(162).

**Localization**

If we look beyond least squares (LS) fitting of Gaussian functions or centroid estimates of candidate spots, we arrive at maximum likelihood estimates (MLE) of the position, i.e. a set of parameters that maximize the likelihood for a particular spot image. The likelihood function is thus a function describing the probability of generating a specific image given a set of parameters, e.g. the fluorophore localization, the point spread function, the noise properties of the camera and possibly the motion blur. Some of the parameters can be measured independently and some are estimated by the MLE optimization. The shape of the likelihood function includes information about the precision of the estimated parameters, e.g. the localization of the molecule that generated the image.

The best average performance of an estimator, expressed as the smallest possible variance of e.g. localization precision, is known as the Cramer-Rao lower bound (CRLB). In the case of the Gaussian point spread function with a relevant noise model, a useful approximate expression was given by Mortensen at al.(163) and further refined by Rieger and Stallinga(164). Strictly, the CRLB is not a statement about the localization precision in a single image; it instead describes the average information content of data generated by a model. In practice, it can be used to estimate uncertainty in single images, although better methods are available. Lindén at al.(165) compared the CRLB to using the peakyness of the likelihood function for the individual spots and found this to be practical in a broader range of conditions. The method was most effective when combined with a Bayesian approach to localization where priors can be used to incorporate physical limitations about the detection system thus stabilize the fitting procedure.

**Linking and trajectory building**

The overall challenge is finding the most likely set of trajectories throughout a movie with dense spots, overlapping trajectories, blinking, bleaching and appearances of new fluorophores. As always, "likely" depends on your assumptions and how complex models you are willing to consider. For example, is it likely that a molecule with a history of fast diffusion keeps moving fast after crossing paths with a stationary molecule or will the encounter change the diffusive properties?

The overall best solution to the problem is arguably to construct all possible trajectories through the spots in all frames and identify the globally most likely set that does not break any constraints and minimizes, for example, the total displacement. This approach is usually impossible due to the vast number of potential trajectories. The u-track algorithm by Jaqaman et al.(166) has become very popular as a way to solve this problem by splitting it into two phases. Trajectory segments are first identified locally with conservative rules such as segment splitting by blinking events. The short segments are then combined into a globally optimal set of trajectories.

Another vital contribution to the tracking problem is the combination of trajectory building and spot identification, which is a beneficial approach when analyzing movies with spots that are blinking and blurred by motion. Sergé et al.(159) developed the multiple-target tracking (MTT) algorithm to link trajectories to an exhaustive set of possible spots in consecutive frames by determining which of the spots that are most likely to belong to a particular trajectory. This strategy allows trajectory properties to influence the identity of the next most probable spot.

## Analysis of trajectories

Inferring intracellular kinetics from single-molecule trajectories requires trajectory property changes as a function of state change. An illustrative example is the change in diffusion rate that occurs when a small labeled molecule binds to a larger structure. To infer the underlying kinetics for a specific set of trajectories, it is usually necessary to assume a number of different dynamic models with different parameters that could potentially have given rise to the data. The Hidden Markov Model (HMM)(Fig. 2B) has been the standard approach for analyzing single-molecule data for decades, including early work on ion channels(167).

However, the use of HMMs is not a magic solution to the model selection problem. To select between a complex model with the associated risk of overfitting and a simple model which does not fit the data, is an obvious challenge. A straightforward MLE will usually fail since it is not possible to directly compare likelihoods obtained from models with different complexity. A typical solution to avoid overfitting is cross-validation by fitting the parameters of models with different complexity to a subset of the data and compare how well they perform on the rest of the dataset(168). An attractive alternative is to use a maximum evidence-based method that automatically penalizes overly complicated models. The idea is that it is unlikely for a complex model to generate data that could be obtained from a simpler model with a smaller parameter space(47).

Recent developments in this area account for the localization inaccuracy in each spot to avoid assigning the molecule to a new diffusive state due to e.g. blurring by out-of-focus diffusion(165). This analysis does not take the localization of each spot for granted, but includes the true trajectory as a hidden variable in the model.

As biology is likely to keep providing us with new intricate kinetics, we cannot expect that all sorts of model complexity can be represented in the HMM. Kinetic rates can for example be different in different parts of space due to the presence of molecules that are not part of the model. When these circumstances are known, however, it is advantageous to include them in the HMM. As an example, Monnier et al. have included directed transports in their analysis since it is essential to model neuronal transport(169).

Importantly, the result of an HMM is not to be interpreted as the "true" description of the intracellular states and kinetics. It is a representation of the high dimensional data squeezed into a low dimensional model that can, at best, help to make sense of the observations. Ideally it should inspire hypotheses that can be tested by other types of experiments.

## Simulations

Not surprisingly, the most general analysis lesson so far is that there are no silver bullets that fit all use cases. Instead, the choice of method will depend on the experimental parameters and the subsequent analysis steps. Similarly, many popular algorithms contain tuning parameters that can be difficult to choose except by trial and error. To get the most out of the experimental data, it is necessary to optimize both the analysis methods and the experimental conditions and to use control experiments tuned to the specific biological process in question. Since ground truth regarding the position and chemical state of single molecules inside living cells is usually impossible to obtain, it is often necessary to resort to simulated data.

Simulations are useful at two different experimental levels. First, we can rarely measure what we want directly and the steps between the parameters that we can model (predict) and the observations are usually too complicated to be analytically tractable. In this case, simulations can help us determine if our conclusions are justified based on the model assumptions. An educational example is presented in(37) where English et al. measure the MSD for the 2D movement of a labeled protein based on a finite number of trajectories, with a finite length distribution and localization accuracy. The slope of the MSD curve differs between cells and levels off at different plateaus, and the diffusion appears slower close to the cell boundary. However, when the experiment is simulated it becomes evident that all data can be explained assuming simple normal diffusion at a single diffusion constant in the confined volume of the cells.

Secondly, we can use simulated data to generate ground truths for evaluation and optimization of spot localization methods and trajectory analysis. In this case, it is important to generate data with more realism than the assumptions used in deriving the analysis model, in order to avoid what is known as the "inverse crime"(170). For example, there is little point in generating test data for spot localization by sampling a Gaussian function on top of a Poisson noise background and use it to test a localization model based on the same assumptions. A better approach would be to sample a measured 3D-PSF and include 3D motion blur and a proper EMCCD noise model to evaluate the expected performance of the method. There are simulation packages available to generate simulated microscopy experiments including stochastic kinetics, cellular compartments, bleaching, blinking, motion blur and noise in the detection system(137, 171).

## Outlook

Single-molecule methods to study intracellular kinetics are likely to revolutionize the way we think about chemistry and physics inside cells. The chemical kinetics in cells is probably unlike the well stirred dilute solutions that we are used to thinking about but it is still unclear how much of the more "exotic" physical chemistry of for example subdiffusion and phase separations matters for understanding the finer details of life(172).

It is however clear that we will need methods to resolve kinetics at the sub-millisecond timescale. Methods like MINFLUX tracking come a long way in this respect; however, macromolecules do not move much at these length scales and we will need other ways than diffusivity to detect binding and conformational states. A desirable development is therefore to combine high-resolution tracking of translational diffusion with other readouts of state such as rotational diffusion, FRET and fluorescence lifetime. Such modalities also push the demand for new properties of dyes and labeling schemes, as well as for statistical learning methods to deal with the new data types.

Another challenge for the future is that of perturbations. One of the apparent advantages of single-molecule observations is that we can learn kinetic information by following the stochastic transitions between states for individual molecules without synchronizing perturbations. However, if we do not know which interaction partners cause the localizations or modulate the rates we do not get very far in connecting the intracellular reaction network. One important direction is therefore to combine single-molecule measurements with specific perturbations to study the impact on intracellular kinetics. To this end, it would be interesting to perform genome-wide screens of single-molecule kinetic measurements, i.e. to study intracellular kinetics in a library of strains with perturbed gene expression(173). Other variations on this theme include studies of mutation libraries to pinpoint which amino acids in the protein of interest are essential for the in vivo interactions.

Finally, data management including storage, annotation, sharing and reanalysis will become an increasingly challenging problem. One may here advocate standardization in data formats and analysis routines to facilitate reproducibility as well as the comparison of results between labs. However, we suspect that the field is still moving too fast to define standards that would not limit the quality of the experiments. The situation for making dynamic measurements in living cells is in this sense different compared to super-resolution imaging of fixed cells, since the measurement scheme and analysis need to adapted much more to the biological process of interest. Concerning kinetic measurements, we are still in a phase where labeling schemes, perturbation methods, acquisition protocols and analysis code need to be tailored to the specific experiment to squeeze out the last bit of information from each photon.

## Acknowledgment

We are grateful to Martin Lindén, Ido Golding, Daniel Jones, David Fange and Magnus Johansson for giving feedback on the text. Work in the Elf lab is supported by grants from the European Research Council (ERC-2013-CoG 616047), the Swedish Research Council (621-2012-4027, 642-2013-7841 and 2016-06213) and the Knut and Alice Wallenberg Foundation (2012.0127 and 2016.0077).